\documentclass[twocolumn,floats,floatfix,nofootinbib,aps,pra]{revtex4-2}

\usepackage{amsfonts,amssymb,amsmath,amsbsy}
\usepackage{physics}
\usepackage{color,calc}
\usepackage{bm}
\usepackage{array}
\usepackage{booktabs}
\usepackage{graphicx}
\usepackage{graphics}
\usepackage{eepic,epsfig}
\usepackage[breaklinks,hidelinks,colorlinks=true,linkcolor=blue,citecolor=blue,filecolor=blue,urlcolor=blue]{hyperref}

\usepackage[dvipsnames]{xcolor}
\definecolor{myblue}{RGB}{0, 0, 255}
\definecolor{darkblue}{rgb}{0,0,0.55}
\definecolor{darkgreen}{rgb}{0,0.39,0}

\usepackage[toc,page]{appendix}

\usepackage{microtype}

\usepackage{tikz}
\usetikzlibrary{arrows.meta, decorations.pathmorphing}

\definecolor{RED}{rgb}{1,0,0}

\def\be{ \begin{equation} }
\def\ee{ \end{equation} }
\def\bea{ \begin{eqnarray} }
\def\eea{ \end{eqnarray} }
\def\bse{ \begin{subequations} }
\def\ese{ \end{subequations} }
\def\ba{ \begin{array} }
\def\ea{ \end{array} }
\def\bt{ \begin{tabular} }
\def\et{ \end{tabular} }

\long\def\/*#1*/{}

\begin{document}

\title{Magnetically Induced Transparency-Absorption and Normal-Anomalous Dispersion Characteristics of ${}^{87}\text{Rb}$ Medium or Any J-Type Configuration Atomic Vapors Subject to a Vector Magnetic Field and a Weak Resonant Pump}

% \title{Magnetically Induced Transparency, Absorption, and Dispersion Characteristics of a ${}^{87}\text{Rb}$ Medium in an J-Type Configuration for Weak Magnetic Field Measurement}
%Magnetically Induced Transparency, Absorption, and Dispersion Characteristics of a ${}^{87}\text{Rb}$ Medium in an J-Type Configuration: A Novel Method for Weak Vector Magnetic Field Measurement
%Magnetic field induced absorption reduction as a tool for vector magnetic field measurement
%https://chatgpt.com/c/68fb3b39-b250-8331-aadf-83f60ff70f32 Lande factors
%https://chatgpt.com/c/68fb7b2b-4074-8328-9baa-ebc0c4e62160 Non-Markovian dynamics
%https://chatgpt.com/c/695b6b4b-f340-8331-bc8e-b358dbfbf7b5
%https://chatgpt.com/c/695bbc5c-8958-8326-a42f-cc30a9088336  

\author{Hayk L. Gevorgyan\textsuperscript{\hyperref[1]{1},\hyperref[2]{2},\hyperref[3]{3},\hyperref[4]{4}}}
\email{hayk.gevorgyan@aanl.am}
\affiliation{
\phantomsection\label{1}{\textsuperscript{1}Quantum Technologies Division, Alikhanyan National Laboratory (Yerevan Physics Institute), 2 Alikhanyan Brothers St., 0036 Yerevan, Armenia}\\
\phantomsection\label{2}{\textsuperscript{2}Experimental Physics Division, Alikhanyan National Laboratory (Yerevan Physics Institute), 2 Alikhanyan Brothers St., 0036 Yerevan, Armenia}\\
\phantomsection\label{3}{\textsuperscript{3}Matinyan Center for Theoretical Physics, Alikhanyan National Laboratory (Yerevan Physics Institute), 2 Alikhanyan Brothers St., 0036 Yerevan, Armenia}\\
\phantomsection\label{4}{\textsuperscript{4}Laboratory of Theoretical Physics, Institute for Physical Research, Armenian National Academy of Sciences, Ashtarak-2, 0203, Armenia}}

% \author{Hayk L. Gevorgyan\textsuperscript{\hyperref[1]{1},\hyperref[2]{2}}}
% \email{hayk.gevorgyan@aanl.am}
% \affiliation{
% \phantomsection\label{1}{\textsuperscript{1}Quantum Technologies Division, Alikhanyan National Laboratory (Yerevan Physics Institute), 2 Alikhanyan Brothers St., 0036 Yerevan, Armenia}\\
% \phantomsection\label{2}{\textsuperscript{2}Experimental Physics Division, Alikhanyan National Laboratory (Yerevan Physics Institute), 2 Alikhanyan Brothers St., 0036 Yerevan, Armenia}}
% \author{Yura P. Malakyan\textsuperscript{\hyperref[3]{3}}}
% \affiliation{\phantomsection\label{3}{\textsuperscript{3}Laboratory of Theoretical Physics, Institute for Physical Research, Armenian National Academy of Sciences, Ashtarak-2, 0203, Armenia}}

%0204
\email{hayk.gevorgyan@aanl.am}

\date{\today }

\begin{abstract}

We have developed an analytical framework for magnetically induced transparency-absorption (MITA) and normal-anomalous dispersion (MINAD) in a weakly driven ${}^{87}\text{Rb}$ vapor, or any J-type three-level system, under a vector magnetic field. By solving the Bloch equations in the stationary, quasi-stationary, and short-pulse regimes, we obtained closed-form expressions for the atomic populations and coherences and identified a bifurcation in the oscillatory dynamics at zero longitudinal Zeeman splitting. The Fourier-domain analysis reveals alternating transparency/absorption and normal/anomalous dispersion with frequency-dependent sign reversals, enabling spectrally selective filtering and group-delay effects. Slow oscillatory behavior in the radio-frequency range makes the system suitable for weak magnetic-field sensing, while fast oscillations at optical frequencies suggest applications in spectral filtering and frequency-comb-like signal shaping. The results provide a theoretical basis for experimental observation of MITA/MINAD and for optimizing atomic-vapor platforms for precision magnetometry and related photonic functionalities.

% We further suggest that the characteristic absorption and dispersion response of very-low-frequency (VLF) probe radio waves interacting with ${}^{87}\text{Rb}$ vapor (or analogous systems) in a J-type configuration, under combined electromagnetic and magnetic driving fields, can be exploited for weak-magnetic-field detections, since the oscillatory behavoir in this case is slow. Furthermore, at higher probe frequencies, where the oscillatory behavoir in fast --- particularly in the visible range—the vapor medium acts as a selective spectral filter, extracting a broadband frequency comb from the optical continuum of incident probe. These results provide a theoretical framework to guide experimental implementation, offering practical strategies for observing MITA/MINAD and optimizing atomic-vapor–based weak magnetic-field sensors.

\end{abstract}

\maketitle

%%%%%%%%%%%%%%%%%%%%%%%%%%%%%%%%%%%%%%%%%%%%%%%%%%%%%%%%%%%%%%%%%%%%%%%%%%%%%%%%%%%%%%%%%%%%%%%%%%%%%%%%%%%%%%%%%%%%%%%%%%%%%%%%%%%%%%%%%

%()%()%()%()%()%()%()%()%()%()%()%
%()%()%()%()%()%()%()%()%()%()%()%
%()%()%()%()%()%()%()%()%()%()%()%

\section{Introduction\label{Sec:intro}}

Resonant nonlinear magneto-optical effects (NMOEs) in atoms have wide applications, including medicine, spectroscopy and precision metrology, geophysics and geology, quantum and optical technologies, fundamental physics \cite{Budker2002}, navigation and space applications \cite{Bloom1962,Budker2000,Hovde2015,Mateos2015}, etc. This research field primarily focuses on two directions: optical pumping and optical probing \cite{Budker2002}. Since we study the influence of light on the medium, the former is the focus of the present paper.

Due to their high sensitivity, optically pumped magnetometers are promising candidates for the detection of weak magnetic fields. Several all-optical techniques have been developed for this purpose \cite{Bloom1962, Budker2000, Dupont-Roc1969}; however, only a few of them enable vector magnetic field measurements \cite{Patton2014, Seltzer2004}, typically requiring the use of two laser fields. The influence of a transverse magnetic field on absorption and fluorescence spectra has attracted growing interest and has been investigated both theoretically and experimentally \cite{Kim2012,Moon2014,Margalit2013a,Margalit2013b,Yudin2010,Cox2011}.

% In \cite{Bloom1962}, the instrument's sensitivity peaks at $\Gamma_r = 3/2 \Gamma_t$, indicating optimal performance with narrow resonance lines and low light intensity. A Bloch-equation-based solution for a spin-1/2 optically pumped system was developed, with the expectation that it could extend to the more complex spin structure of Rb atoms. 

In Refs. \cite{Kim2012,Moon2014}, the influence of a magnetic field on laser field absorption is investigated experimentally, where a reduction in absorption is referred to as magnetic-field-induced transparency (MIT) --- in analogy with electromagnetically induced transparency (EIT) \cite{Fleischhauer2005, Boller1991}. However, there is a fundamental difference between MIT and EIT: in MIT, transparency arises from atomic population redistribution in the ground state caused by a transverse magnetic field, whereas EIT results from destructive interference between two absorption pathways of the probe field. %The underlying mechanism of MIT is elucidated and supported by our theoretical calculations.

In the present paper, we investigate the \emph{weak excitation} or \emph{weak light regime} of magnetic-field-assisted optical pumping of ${^{87}}\text{Rb}$ atoms. Both transverse and longitudinal magnetic field components are considered, i.e., a vector magnetic field. In this regime, the Bloch equations decouple, allowing for an analytical solution. For simplicity, we assume resonant excitation, although an analytical solution in the off-resonant case also appears feasible; this will be addressed in future work and compared with the results presented here. We address separated three-level system with schematic J-type configuration\footnote{It resembles the letter ``J.''} Fig.~\ref{fig:model} of relevant atomic levels, where $\ket{1}$ and $\ket{2}$ are Zeeman sublevels within the same ground hyperfine state with the same total angular momentum $F$ but different magnetic quantum numbers $m_F$ and $m_F + 1$. $\ket{3}$ is a sublevel of an excited hyperfine state with total angular momentum $F' = F+1$. As usual, a right-circularly polarized light, being carrier of photon angular momentum $\hbar$ per photon, drives $\Delta m = 1$ transition from $\ket{2}$ to $\ket{3}$. For experimental purpose, an example of $^{87}\text{Rb}$ $D_2$ line cyclic transition can be numerically analyzed \cite{SteckData}. The analytical solution of Bloch equations provides ultimate guide for experimentalists in study of behavior of coherences and populations of the system in \emph{stationary}, \emph{quasi-stationary}, and \emph{short-pulse regimes}. Moreover, observation of light absorption coefficient using expressions for $\mathfrak{Im}({\rho}_{32})$, one can know how to measure magnetic fields, even weak ones. Likewise, for certain parameters $\Omega, \Delta_L, \Delta_2, L_g, \Gamma$, \emph{magnetically-induced transparency} can be detected.   

\section{Theoretical model}
\begin{figure}[ht]
    \centering
\begin{tikzpicture}[scale=1.50]
    \draw (-0.75,1.55) -- (0.75,1.55) -- (0.75,1.45) -- (-0.75,1.45) -- cycle;
    \fill[black] (-0.75,1.55) -- (0.75,1.55) -- (0.75,1.45) -- (-0.75,1.45) -- cycle node [midway, left] {$\ket{3}$};
    \draw (-0.75,-1.45) -- (0.75,-1.45) -- (0.75,-1.55) -- (-0.75,-1.55) -- cycle;
    \fill[black] (-0.75,-1.45) -- (0.75,-1.45) -- (0.75,-1.55) -- (-0.75,-1.55) -- cycle node [midway, left] {$\ket{2}$};
    \fill[black] (-0.75-1.3,-1.45-0.5) -- (0.75-1.3,-1.45-0.5) -- (0.75-1.3,-1.55-0.5) -- (-0.75-1.3,-1.55-0.5) -- cycle node [midway, left] {$\ket{1}$};
    \draw[dashed] (-0.75,1) -- (0.75,1);
    \draw[red][-Stealth][ultra thick] (0,-1.5) -- (0,1) node [midway, left] {$\Omega(t), \sigma^+$}; 
    \draw[blue][Stealth-Stealth][ultra thick] (0,1) -- (0,1.5) node [midway, right] {$\Delta_L$};
    \draw[blue][Stealth-Stealth][ultra thick] (-0.65,-1.5-0.5) -- (-0.65,-1.5) node [midway, right] {$\Delta_2$};
    \coordinate (A) at (0,-1.5);
    \coordinate (B) at (-1.3,-1.5-0.5);
    \draw[<->, ultra thick, darkgreen] (A) .. controls (-0.3,-2.5) and (-1,-2.5) .. (B) node[pos=0.2, right] {$L_g$};
\end{tikzpicture}
\caption{Schematic configuration}
    \label{fig:model}
\end{figure}
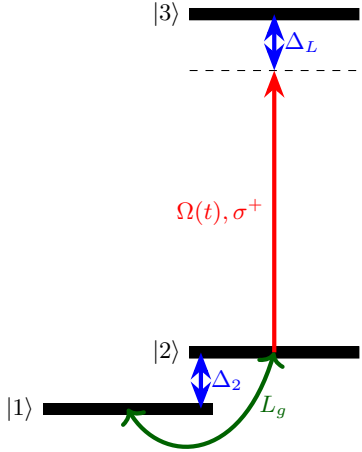

In the context of the rotating wave approximation (RWA), the interaction Hamiltonian of a three-level system subjected to both optical and magnetic fields, as illustrated in Figure~\eqref{fig:model}, takes the following form in the rotating frame:
\be\label{Hamiltonian}
H = \hbar \Delta_L \sigma_{33} + \hbar \Delta_2 \sigma_{22} - \hbar (\Omega \sigma_{32} + L_g \sigma_{21} + H.c.), 
\ee
where $\Delta_L \overset{\Delta}{=} \omega_{32} - \omega_L$ and $\Omega$ represent the detuning and Rabi frequency of the optical field, respectively, which interacts with the transition  $\ket{2} \rightarrow \ket{3}$. The term $\Delta_2 = \mu_0 g B_z / \hbar$ corresponds to the Zeeman effect caused by the longitudinal magnetic field $B_z$, reflecting the energy difference between the split states $\ket{1}$ and $\ket{2}$. Additionally, $L_g=\mu_0 g B_x / \hbar$ is the Larmor frequency associated with the transverse component of the magnetic field $B_x$, which can induce transitions between the states $\ket{1} \leftrightarrow \ket{2}$.

According to the Liouville-von Neumann equation
\be\label{Liouville}
\dot{\rho} = \frac{i}{\hbar} [\rho, H] - \frac{1}{2} \{\Gamma, \rho\} + \Lambda,
\ee
the time evolution of the density matrix elements $\rho_{ij}$ of a quantum system \eqref{Hamiltonian} is given by the optical Bloch equations for the population 
\bse\label{population}
\begin{gather}
\dot{\rho}_{11} = i (L^*_g \rho_{21} - L_g \rho_{12}) + \Gamma_{2} \rho_{22} + \Gamma_{31} \rho_{33} - \Gamma_{2} \rho_{11}, \\
\dot{\rho}_{22} = i (L_g \rho_{12} - L^*_g \rho_{21} + \Omega^* \rho_{32} - \Omega \rho_{23}) - \Gamma_{2} \rho_{22} + \notag\\
+ \Gamma_{32} \rho_{33} - \Gamma_{2} \rho_{11}, \\
\dot{\rho}_{33} = i (\Omega \rho_{23} - \Omega^* \rho_{32}) - \Gamma_3 \rho_{33}, \label{poprho33}
\end{gather}
\ese
and the coherence
\bse\label{coherence}
\begin{gather}
\dot{\rho}_{21} =  (- i \Delta_2 - \gamma_{21}) \rho_{21} - i L_g (\rho_{22} - \rho_{11}) + i \Omega^* \rho_{31}, \\ 
\dot{\rho}_{31} =  (- i \Delta_L - \gamma_{31}) \rho_{31} + i \Omega \rho_{21} - i L_g \rho_{32}, \\
\dot{\rho}_{32} =  (- i (\Delta_L - \Delta_2) - \gamma_{32}) \rho_{32} + i \Omega \rho_{22} - \notag\\
- i L^*_g \rho_{31} - i \Omega \rho_{33},
\end{gather}
\ese
dynamics, where $\Gamma_3 \overset{\Delta}{=} \Gamma_{31} + \Gamma_{32}$ is the total decay rate of the excited state (one may take $\Gamma_{31} = \Gamma_{32} = \Gamma_3 /2 \overset{\Delta}{=} \Gamma$), $\Gamma_2 \overset{\Delta}{=} \Gamma_{21}$ and decoherence rates are always equal $\gamma_{31} = \gamma_{32} = \Gamma$.

Let's consider that $\Gamma_2 = \gamma_{21} = \Delta_L \approx 0$, the optical and the magnetic fields are in phase $\Omega^* = \Omega$, $L^*_g = L_g$; and the system is in a \emph{weak excitation regime}: $\rho_{33}, \rho_{31}, \rho_{32} \approx 0$. The latter leads to the case $\dot{\rho}_{33} \approx 0$ and $n_1 + n_2 \approx 0$, where $\rho_{11} (t = -\infty) \overset{\Delta}{=} n_1$ and $\rho_{22} (t = -\infty) \overset{\Delta}{=} n_2$ are the initial populations of states $\ket{1}$ and $\ket{2}$, respectively. The optical Bloch equations \eqref{population} and \eqref{coherence} are now presented in their final form
for the dynamics of populations
\bse\label{population1}
\begin{gather}
\dot{\rho}_{11} = - 2 L_g \mathfrak{Im}(\rho_{21}), \\
\dot{\rho}_{21} = - i \Delta_2 \rho_{21} - i L_g (1 - 2 \rho_{11}),
\end{gather}
\ese
and coherences
\bse\label{coherence1}
\begin{gather}
\dot{\rho}_{31} =  - \Gamma \rho_{31} - i L_g \rho_{32} + i \Omega \rho_{21}, \\
\dot{\rho}_{32} =  - i L_g \rho_{31} + (i \Delta_2 - \Gamma) \rho_{32} + i \Omega \rho_{22}.
\end{gather}
\ese

\section{Derivation method}
Here, we use the following derivation method
\begin{enumerate}
\item Eqs.~\eqref{population1} and~\eqref{coherence1} are reformulated into the system of equations~\eqref{population2} and~\eqref{coherence2} with the real variables:
\bse\label{population2}
\begin{align}
& \dot{u}_1 = -2 L_g w_1, \\
& \dot{v}_1 = \Delta_2 \, w_1, \\
& \dot{w}_1 = 2 L_g u_1 - \Delta_2 \, v_1 - L_g, \label{population2_3}
\end{align}
\ese 
where $u_1 \overset{\Delta}{=} {\rho}_{11}$, $v_1 \overset{\Delta}{=} \mathfrak{Re}({\rho}_{21})$, $w_1 \overset{\Delta}{=} \mathfrak{Im}({\rho}_{21})$, 
\bse\label{coherence2}
\begin{align}
& \dot{u}_2 = - \Gamma u_2 + L_g m_2 - \Omega \, w_1, \\
& \dot{v}_2 = - \Gamma v_2 - L_g w_2 + \Omega \, v_1, \\
& \dot{w}_2 = L_g v_2 - \Gamma w_2 - \Delta_2 \, m_2 , \\
& \dot{m}_2 = - L_g u_2 + \Delta_2 \, w_2 - \Gamma m_2 + \Omega \, (1-u_1),
\end{align}
\ese 
where $u_2 \overset{\Delta}{=} \mathfrak{Re}({\rho}_{31})$, $v_2 \overset{\Delta}{=} \mathfrak{Im}({\rho}_{31})$, $w_2 \overset{\Delta}{=} \mathfrak{Re}({\rho}_{32})$, $m_2 \overset{\Delta}{=} \mathfrak{Im}({\rho}_{32})$.

\item Since Eq.~\eqref{population2} is full-fledged in the sense that it forms a closed system of equations for the variables $u_1$, $v_1$ and $w_1$, it can be solved separately from Eq.~\eqref{coherence2}. Hence, Eqs.~\eqref{population2} and~\eqref{coherence2} are solved separately treated as two separate matrix differential equations. The first, inhomogeneous matrix differential equation with time independent inhomogeneity ($-L_g$ in Subeq.~\eqref{population2_3}) is solved using the eigenvector method and the variation of constants. The second, inhomogeneous matrix differential equation is solved using the eigenvector method for it's homogeneous part and is extended for the whole matrix differential equation using the variation of constants to account for the time-dependent inhomogeneity due to ``unknown'' time dependences $u_1 (t)$, $v_1(t)$, $w_1 (t)$, i.e., $\rho_{11} (t)$, $\rho_{21} (t)$, and $\Omega(t)$ in the case of short-pulse regime (see Sec.~\ref{Subsec:short-pulse}). Then, actually known dependences $u_1 (t)$, $v_1(t)$, $w_1 (t)$, solutions from Eq.~\eqref{population2}, are placed in the solved dynamics of \eqref{coherence2}. 
\end{enumerate}

\section{Solution for populations}
A solution of Eq.~\eqref{population2} gives
\bse\label{solution_population}
\begin{gather}
{\rho}_{11} = \frac{2 L^2_g}{\alpha^2} + \frac{\Delta_2}{2 L_g} f_1 - \frac{2 L_g}{\alpha} \left(f_2 \sin{\alpha t} + f_3 \cos{\alpha t} \right), \\ 
{\rho}_{21} = - \frac{\Delta L_g}{\alpha^2} + f_1 + \frac{\Delta_2}{\alpha} f_2 \sin{\alpha t}  + \frac{\Delta_2}{\alpha} f_3 \cos{\alpha t} + \notag \\
+ i \left[f_2 \cos{\alpha t} - f_3 \sin{\alpha t} \right], \\
f_1 = \frac{4 L^2_g}{\alpha^2} \mathfrak{Re}({\rho}_{21} (0)) + \frac{2 \Delta_2 L_g}{\alpha^2} \rho_{11}(0), \notag \\
f_2 = \mathfrak{Im}({\rho}_{21} (0)), \notag \\
f_3 = \frac{L_g}{\alpha} + \frac{\Delta_2}{\alpha} \mathfrak{Re}({\rho}_{21}(0)) - \frac{2 L_g}{\alpha} \rho_{11}(0). \notag 
\end{gather}
\ese 

For the initial conditions $\rho_{11} (0) = 1$ and $\rho_{21} (0) = 0$, we get that $f_1 = \frac{2 \Delta_2 L_g}{\alpha^2}$, $f_2 = 0$, $f_3 = - \frac{L_g}{\alpha}$ and the solution has the following form
\bse\label{solution_population_initial}
\begin{align}
& {\rho}_{11} = 1 - \frac{2 L^2_g}{\alpha^2} + \frac{2 L^2_g}{\alpha^2} \cos{\alpha t}, \\ 
& {\rho}_{21} = \frac{\Delta_2 L_g}{\alpha^2} - \frac{\Delta_2 L_g}{\alpha^2} \cos{\alpha t} + i \frac{L_g}{\alpha} \sin{\alpha t}, 
\end{align}
\ese 
where $\alpha \overset{\Delta}{=} \sqrt{4 L^2_g + \Delta^2_2}$.

\section{Solution for coherences}
In the matrix form, Eq.~\eqref{coherence2} reads as 
\be\label{coherences_matrix}
\begin{gathered}
\dot{X} = A X + B(t),\\
A = \left[ \begin{array}{cccc} - \Gamma & 0 & 0 & L_g \\ 0 & - \Gamma & -L_g & 0 \\ 0 & L_g & - \Gamma & - \Delta_2 \\ - L_g & 0 & \Delta_2 & - \Gamma \end{array} \right], 
\quad B(t) = \left[ \begin{array}{cccc} - \Omega(t) \mathfrak{Im}({\rho}_{21}(t)) \\ \Omega(t) \mathfrak{Re}({\rho}_{21}(t)) \\ 0 \\ \Omega(t) {\rho}_{22}(t) \end{array} \right].
\end{gathered}
\ee

\subsection{Case of $\Delta_2 = 0$}

Eigenvector method gives

\be\label{Xd}
\begin{gathered}
X = \left[ \begin{array}{cccc} \mathfrak{Re}({\rho}_{31}) \\ \mathfrak{Im}({\rho}_{31}) \\ \mathfrak{Re}({\rho}_{32}) \\ \mathfrak{Im}({\rho}_{32}) \end{array} \right] = e^{- \Gamma t} \left[ \begin{array}{cccc}  d_1(t) \sin{L_g t} + d_3(t) \cos{L_g t} \\ d_2(t) \sin{L_g t} + d_4(t) \cos{L_g t} \\ d_4(t) \sin{L_g t} - d_2(t) \cos{L_g t} \\ - d_3(t) \sin{L_g t} + d_1(t) \cos{L_g t}  \end{array} \right]  \\
= e^{- \Gamma t} \left[ \begin{array}{cccc}  a_1(t) \sin{\left(L_g t + \phi_1 (t)\right)} \\ a_2(t) \sin{\left(L_g t + \phi_2 (t)\right)} \\ - a_2(t) \cos{\left(L_g t + \phi_2 (t)\right)} \\ a_1(t) \cos{\left(L_g t + \phi_1 (t)\right)}  \end{array} \right],
\end{gathered}
\ee
\begin{widetext}
Variation of coefficients (constants) gives

\bse\label{Xdvar}
\begin{align}
& d_1(t) = \int_{- \infty}^{t} e^{\Gamma \tau} \Omega(\tau) \left({\rho}_{22}(\tau) \cos{L_g \tau} - \mathfrak{Im}({\rho}_{21}(\tau)) \sin{L_g \tau} \right) \,d\tau + d'_1, \\
& d_3(t) = - \int_{- \infty}^{t} e^{\Gamma \tau} \Omega(\tau) \left({\rho}_{22}(\tau) \sin{L_g \tau} + \mathfrak{Im}({\rho}_{21}(\tau)) \cos{L_g \tau} \right) \,d\tau + d'_3, \\
& d_2(t) = \int_{- \infty}^{t} e^{\Gamma \tau} \Omega(\tau) \mathfrak{Re}({\rho}_{21}(\tau)) \sin{L_g \tau} \,d\tau + d'_2, \\
& d_4(t) = \int_{- \infty}^{t} e^{\Gamma \tau} \Omega(\tau) \mathfrak{Re}({\rho}_{21}(\tau)) \cos{L_g \tau} \,d\tau + d'_4, \\
& a_1 (t) = \sqrt{d^2_1 (t) + d^2_3 (t)}, \quad a_2 (t) = \sqrt{d^2_2 (t) + d^2_4 (t)}, \quad \tan{\phi_1 (t)} = \frac{d_3 (t)}{d_1 (t)}, \quad \tan{\phi_2 (t)} = \frac{d_4 (t)}{d_2 (t)}.
\end{align}
\ese 

\subsection{Case of $\Delta_2 \neq 0$}
Eigenvector method gives

\be\label{Xc}
\begin{gathered}
X = \left[ \begin{array}{cccc} \mathfrak{Re}({\rho}_{31}) \\ \mathfrak{Im}({\rho}_{31}) \\ \mathfrak{Re}({\rho}_{32}) \\ \mathfrak{Im}({\rho}_{32}) \end{array} \right] = e^{- \Gamma t} \left[ \begin{array}{cccc}  c_1(t) \cos{\varkappa_+ t} - c_2(t) \sin{\varkappa_+ t} + c_3(t) \cos{\varkappa_- t} - c_4(t) \sin{\varkappa_- t} \\ -\frac{L_g^2 \omega_+}{\Delta_2 \varkappa_+} \left( c_2(t) \cos{\varkappa_+ t} + c_1(t) \sin{\varkappa_+ t} \right) - \frac{L_g^2 \omega_-}{\Delta_2 \varkappa_-} \left( c_4(t) \cos{\varkappa_- t} + c_3(t) \sin{\varkappa_- t} \right) \\ \frac{L_g \omega_+}{\Delta_2} \left( c_1(t) \cos{\varkappa_+ t} - c_2(t) \sin{\varkappa_+ t} \right) + \frac{L_g \omega_-}{\Delta_2} \left( c_3(t) \cos{\varkappa_- t} - c_4(t) \sin{\varkappa_- t} \right) \\ - \frac{\varkappa_+}{L_g} \left( c_2(t) \cos{\varkappa_+ t} + c_1(t) \sin{\varkappa_+ t} \right) - \frac{\varkappa_-}{L_g} \left( c_4(t) \cos{\varkappa_- t} + c_3(t) \sin{\varkappa_- t} \right) \end{array} \right] \\
= e^{- \Gamma t} \left[ \begin{array}{cccc}  A_1 (t) \cos{(\varkappa_+ t + \varphi_1 (t))} + A_2 (t) \cos{(\varkappa_- t + \varphi_2 (t))} \\
- \frac{L_g^2 \omega_+}{\Delta_2 \varkappa_+} A_1 (t) \sin{(\varkappa_+ t + \varphi_1 (t))} - \frac{L_g^2 \omega_-}{\Delta_2 \varkappa_-} A_2 (t) \sin{(\varkappa_- t + \varphi_2 (t))} \\
\frac{L_g \omega_+}{\Delta_2} A_1 (t) \cos{(\varkappa_+ t + \varphi_1 (t))} + \frac{L_g \omega_-}{\Delta_2} A_2 (t) \cos{(\varkappa_- t + \varphi_2 (t))} \\
- \frac{L_g (1 - \omega_+)}{\varkappa_+} A_1 (t) \sin{(\varkappa_+ t + \varphi_1 (t))} - \frac{L_g (1 - \omega_-)}{\varkappa_-} A_2 (t) \sin{(\varkappa_- t + \varphi_2 (t))} 
\end{array} \right].
\end{gathered}
\ee
Variation of coefficients (constants) gives

\bse\label{Xcvar}
\begin{align}
& c_1(t) = \int_{- \infty}^{t} g(\tau) \cos{\varkappa_+ \tau} + k(\tau) \sin{\varkappa_+ \tau} \,d\tau + c'_1, \quad f(t) = \frac{\omega_+}{\xi} \Omega(t) \mathfrak{Im}(\rho_{21}(t)) e^{\Gamma t},\\
& c_2(t) = \int_{- \infty}^{t} k(\tau) \cos{\varkappa_+ \tau} - g(\tau) \sin{\varkappa_+ \tau} \,d\tau + c'_2, \quad g(t) = - \frac{\omega_-}{\xi} \Omega(t) \mathfrak{Im}(\rho_{21}(t)) e^{\Gamma t}, \\
& c_3(t) = \int_{- \infty}^{t} f(\tau) \cos{\varkappa_- \tau} + l(\tau) \sin{\varkappa_- \tau} \,d\tau + c'_3, \quad k(t) = \frac{\varkappa_+}{\xi L_g} \Omega(t) e^{\Gamma t} \left( \frac{\Delta_2}{L_g} (1 - \omega_-) \mathfrak{Re}(\rho_{21}(t)) - \omega_- \rho_{22}(t) \right), \\
& c_4(t) = \int_{- \infty}^{t} l(\tau) \cos{\varkappa_- \tau} - f(\tau) \sin{\varkappa_- \tau} \,d\tau + c'_4, \quad l(t) = \frac{\varkappa_-}{\xi L_g} \Omega(t) e^{\Gamma t} \left( - \frac{\Delta_2}{L_g} (1 - \omega_+) \mathfrak{Re}(\rho_{21}(t)) + \omega_+ \rho_{22}(t) \right),  \\
& A_1 (t) = \sqrt{c^2_1 (t) + c^2_2 (t)}, \quad A_2 (t) = \sqrt{c^2_3 (t) + c^2_4 (t)}, \quad \tan{\varphi_1 (t)} = \frac{c_2 (t)}{c_1 (t)}, \quad \tan{\varphi_2 (t)} = \frac{c_4 (t)}{c_3 (t)},
\end{align}
\ese 
where the parameters are and the following relations exist
\be
\begin{gathered}
\varkappa_\pm \overset{\Delta}{=}  \sqrt{L_g^2 + \frac{\Delta_2^2}{2}
\pm \frac{\Delta_2^2}{2} \sqrt{1 + \frac{4 L_g^2}{\Delta_2^2}}}, \quad \omega_\pm \overset{\Delta}{=}  1 / \left( \frac{1}{2} \mp \sqrt{\frac{1}{4} + \frac{L_g^2}{\Delta_2^2}} \right), \quad \xi \overset{\Delta}{=}  \omega_- - \omega_+ = \frac{2 \Delta_2}{L_g} \sqrt{\frac{\Delta_2^2}{4 L_g^2} + 1}, \quad \varkappa_\pm^2 = L_g^2 (1 - \omega_\pm), \\
L_g^2 + \frac{\Delta_2^2}{2} > \frac{\Delta_2^2}{2} \sqrt{1 + \frac{4 L_g^2}{\Delta_2^2}} ~~~ \text{as}  ~~~ L_g \neq 0; \quad \lvert\omega_+\rvert = - \omega_+, ~~~ \omega_+ < 0, ~~~ \omega_- > 0; \quad \varkappa_+ \approx \varkappa_- \approx L_g ~~~ \text{when} ~~~ \Delta_2 \rightarrow 0. 
%& \eta_\pm^2 \overset{\Delta}{=} L^2 - \varkappa_\pm^2,
\end{gathered}
\ee

\end{widetext}

A bifurcation occurs when a small change in a parameter leads to a qualitative shift in the system's dynamics. In the present case, as $\Delta_2$ transitions from zero, the system undergoes a transition from a single-mode oscillatory behavior to a two-mode regime. Specifically, the system, initially exhibiting damped oscillations (with time dependent amplitudes $d_1 (t)$, $d_2 (t)$, $d_3 (t)$, $d_4 (t)$) with a single frequency $L_g$, changes to damped oscillations (with time dependent amplitudes $c_1 (t)$, $c_2 (t)$, $c_3 (t)$, $c_4 (t)$) with two distinct frequencies 
$\varkappa_-$ and $\varkappa_+$. Therefore, $\Delta_2 = 0$ marks a bifurcation point, where the system's dynamical behavior alters fundamentally. Due to the time-dependent nature of the coefficients (amplitudes), arising also from the time dependence of the internal factors $\rho_{11} (t)$ and $\rho_{21} (t)$, and the external driving term $\Omega (t)$, the system exhibits more complex behavior than the conventional damped oscillations, but $\Delta_2$ is still a bifurcation point. 

Using Eq.~\eqref{population} and $\mathfrak{Im}({\rho}_{32})$ from Eq.~\eqref{Xc}, let's find an evaluating formula for $\rho_{33}$, although it is approximated to zero. The Eq.~\eqref{poprho33} reads 
\be
\dot{\rho}_{33} = - 2 \Gamma \rho_{33} + 2 \Omega \rho_{33},
\ee
which gives the solution
\be\label{rho33sol}
\begin{gathered}
\rho_{33} = s(t) e^{- 2 \Gamma t}, \\
s(t) = \int_{- \infty}^{t} 2 \Omega(\tau) e^{2 \Gamma \tau} \mathfrak{Im}({\rho}_{32}) \,d\tau + s',
\end{gathered}
\ee
where, in case of $\Delta_2 = 0$, $\mathfrak{Im}({\rho}_{32})$ is given by Eq.~\eqref{Xd}, and in general case $\Delta_2 \neq 0$ by Eq.~\eqref{Xc}. The integral \eqref{rho33sol} is a causal convolution $(F * K)(t) = \int_{- \infty}^{t} F(\tau) G(t-\tau) \,d\tau = \int_{- \infty}^{\infty} F(\tau) G(t-\tau) \Theta(t - \tau) \,d\tau$ of $F(t) = 2 \Omega \mathfrak{Im}({\rho}_{32}$ with the exponential kernel $K(t) = e^{-2\Gamma t}$, where $\Theta(t)$ is a Heaviside step function. Causal convolutions often seen in physics and engineering, and so in our case for $\rho_{31}$ and $\rho_{32}$ (see Eqs.~\eqref{Xd} and~\eqref{Xc}). Note, that $\rho_{33}$ is a causal convolution of expression composed of $\mathfrak{Im}({\rho}_{32})$, which is itself a causal convolution of expressions $A_1 (t)$ and $A_2 (t)$, each composed of ${\rho}_{21}$ and ${\rho}_{22}$. Note also, that the found expressions are valid only in a weak excitation regime $\Omega \ll \Gamma$, since, in the general case, the equations are no more separated, and it would be necessary to solve the system of coupled equations, Eqs.~\eqref{population} and~\eqref{coherence}, numerically.
 %$\Omega \ll \min{(\Delta_2, L_g, \Gamma)}$

\subsection{Stationary regime}
In the stationary regime: $\Gamma T \gg 1$ and $\gamma_0 T > 1$ and $\Omega = const$ ($\dot{\rho}_{11}=\dot{\rho}_{21}=\dot{\rho}_{31}=\dot{\rho}_{32}=0$ and $\gamma_0 \neq 0$), one gets
\be\label{ResultsStationary1}
\rho_{21} = \frac{L_g}{\Delta_2} \left(2 \rho_{11} - 1\right)
\ee
and 
\bse\label{ResultsStationary2}
\begin{align}
\rho_{31} &= \frac{i \Omega}{\Gamma (\Gamma - i \Delta_2) + L_g^2} \left((\Gamma - i \Delta_2) \rho_{21} - i L_g \rho_{22} \right), \\
\rho_{32} &= \frac{i \Omega}{\Gamma (\Gamma - i \Delta_2) + L_g^2} \left( - i L_g \rho_{21} + \Gamma \rho_{22} \right).
\end{align}
\ese
or
\begin{widetext}
\bse
\begin{align}
\mathfrak{Re}({\rho}_{31}) &= \frac{\Omega}{\beta(\Gamma, \Delta_2, L_g)} \left(\Delta_2 L_g^2 \mathfrak{Re}({\rho}_{21}) - \Gamma (\Gamma^2 + L_g^2 + \Delta^2_2) \mathfrak{Im}({\rho}_{21}) + L_g (\Gamma^2 + L_g^2) \rho_{22} \right), \notag \\
\mathfrak{Im}({\rho}_{31}) &= \frac{\Omega}{\beta(\Gamma, \Delta_2, L_g)} \left(\Gamma (\Gamma^2 + L_g^2 + \Delta^2_2) \mathfrak{Re}({\rho}_{21}) + \Delta_2 L_g^2 \mathfrak{Im}({\rho}_{21}) + \Gamma \Delta_2 L_g \rho_{22} \right), \notag \\
\mathfrak{Re}({\rho}_{32}) &= \frac{\Omega}{\beta(\Gamma, \Delta_2, L_g)} \left(L_g (\Gamma^2 + L^2_g) \mathfrak{Re}({\rho}_{21}) - \Gamma \Delta_2 L_g \mathfrak{Im}({\rho}_{21}) - \Gamma^2 \Delta_2 \rho_{22} \right), \notag \\
\mathfrak{Im}({\rho}_{32}) &= \frac{\Omega}{\beta(\Gamma, \Delta_2, L_g)} \left(L_g (\Gamma^2 + L^2_g) \mathfrak{Im}({\rho}_{21}) + \Gamma \Delta_2 L_g \mathfrak{Re}({\rho}_{21}) + \Gamma (\Gamma^2 + L^2_g) \rho_{22} \right), \notag \\
& \beta(\Gamma, \Delta_2, L_g) = (\Gamma^2 + L^2_g)^2 + \Gamma^2 \Delta^2_2.
\notag
\end{align}
\ese
\end{widetext}

The Eq.~\eqref{ResultsStationary1} is valid for any $\Delta_2$ and $L_g$, since always $\kappa = \frac{\abs{\Delta_2}}{2\sqrt{L_g^2 + \Delta_2^2/4}} \leq 1$ due to Cauchy-Schwartz inequality $\abs{\rho_{21}}^2 \leq \rho_{11} \rho_{22}$, and, accordingly, $\rho_{11}$ must be within the following interval 

\be
\rho_{11} \in \left[\frac{1 - \kappa}{2}, \quad \frac{1 + \kappa}{2}\right].
\ee

Although $\rho_{11} + \rho_{22} \approx 1$, there is also an additional counter-constraint (to its neglect) on $\rho_{33}$ imposed by the Cauchy–Schwarz inequalities $\abs{\rho_{31}}^2 \leq \rho_{33} \rho_{11}$ and $\abs{\rho_{32}}^2 \leq \rho_{33} \rho_{22}$:

\be\label{SCrho33}
\begin{gathered}
\rho_{33} \geq \max\left\{{\frac{\mathfrak{Re}^2({\rho}_{31}) + \mathfrak{Im}^2({\rho}_{31})}{\rho_{11}}, \frac{\mathfrak{Re}^2({\rho}_{32}) + \mathfrak{Im}^2({\rho}_{32})}{\rho_{22}}}\right\}, \\
\rho_{33} \ll \min\left\{\rho_{11}, 1 - \rho_{11}\right\}.
\end{gathered}
\ee

\subsection{Quasi-stationary regime}

In the quasi-stationary regime: $\Gamma T \gg 1$ and $\gamma_0 = 0$ and $\Omega = \Omega (t)$ ($\dot{\rho}_{31}=\dot{\rho}_{32}=0$), one gets the following --- $\rho_{11}(t)$ and $\rho_{22}(t)$ are subjected to the \eqref{solution_population} dynamics, and $\rho_{31}$ and $\rho_{32}$ linearly depend on them as in \eqref{ResultsStationary2}, but populations and a resonant pump are time-dependent.  

The corresponding formulas are valid any $\Delta_2$, $L_g$ and at any time $t$, for which Cauchy-Schwartz inequality $\abs{\rho_{21} (t)}^2 \leq \rho_{11} (t) \rho_{22} (t)$ is true, and, accordingly, the function $\rho_{11} (t)$ must satisfy the following inequalities
\bse\label{CStimedep}
\begin{align}
& \rho^2_{11} + \frac{y-1}{1+x} \rho_{11} + \frac{z + \dot{\rho}^2_{11} / 4 L_g^2}{1 + x} \leq 0, \\
& x = \frac{\Delta_2^2}{4 L^2_g}, \quad y = - \frac{\Delta_2}{L_g} \frac{\Delta_2^2 + 4 L^2_g}{4 L^2_g}, \quad z = \frac{(\Delta_2^2 + 4 L_g^2)^2}{16 L_g^2}, \notag \\
& \mathfrak{Re}({\rho}_{21}) = - \frac{\Delta_2}{2 L_g} \rho_{11} + \frac{\Delta_2^2 + 4 L_g^2}{4 L_g^2} f_1, \\
& \mathfrak{Im}({\rho}_{21}) = - \frac{\dot{\rho}_{11}}{2 L_g}
\end{align}
\ese
hence, according to the latter formulas, the corresponding inequalities for $\mathfrak{Re}({\rho}_{21})$ and $\mathfrak{Im}({\rho}_{21})$ can be derived.

Eq.~\eqref{SCrho33} must hold too but now for a time-dependent dynamics.

\subsection{Short-pulse regime}\label{Subsec:short-pulse}

In the short-pulse regime: $\Gamma T \ll 1$ and $\Omega = \Omega (t)$ ($\Gamma = \gamma_0 = 0$), the dynamics of $\rho_{11} (t)$ and $\rho_{21} (t)$ are governed by the law \eqref{solution_population}, when the dynamics of $\rho_{31}$ and $\rho_{32}$:
\begin{itemize}
\item by the law \eqref{Xd} in the case of no longitudinal component of a magnetic field or negligible splitting $\Delta_2 = 0$, or
\item by the law \eqref{Xc} in the general case $\Delta_2 \neq 0$.
\end{itemize}  

Again, Cauchy-Schwartz time-dependent inequalities \eqref{CStimedep} and \eqref{SCrho33} must hold.

\section{Absorption and Dispersion}

Absorption coefficient and dispersion (for dilute vapors $n_0 \approx 1$): 
%https://chatgpt.com/c/6899cb62-b718-8326-9488-5bd79457dca3
\bse
\begin{align}
\alpha (\omega) &= \frac{4\pi \omega N \mu_{32}}{c} \mathfrak{Im}({\rho}_{32} (\omega)), \\
\Delta n(\omega) &= 2\pi N \mu_{32} \mathfrak{Re}({\rho}_{32} (\omega)), 
\end{align}
\ese
where $\mu_{32}$ is a dipole matrix element of the optical transition (esu $\cdot$ cm), $N$ is the atomic number density of the vapor (cm$^{-3}$), $\omega$ is the angular frequency of the incident probe electromagnetic field (rad/s), $c = 3 \cdot 10^{10}$ cm/s is a light speed in a vacuum, and real and imaginary parts of $\rho_{32}$ coherence are Fourier-transformed:

\bse
\begin{align}
\mathfrak{Im}({\rho}_{32} (\omega)) &= \mathfrak{Re} [\mathcal{FT} \left(\mathfrak{Im}({\rho}_{32} (t)\right)], \\
\mathfrak{Re}({\rho}_{32} (\omega)) &= \mathfrak{Re} [\mathcal{FT} \left(\mathfrak{Re}({\rho}_{32} (t)\right)], \\
\mathcal{FT} (Z (t)) &= \int\limits_{- \infty}^{\infty} Z(t) \, exp{(- i \omega t)} \, dt. \notag 
\end{align}
\ese

\section{Weak magnetic field measurement using VLF radio-wave probes}

\begin{figure}[t]
\bt{r}
\centerline{\includegraphics[width=1\columnwidth]{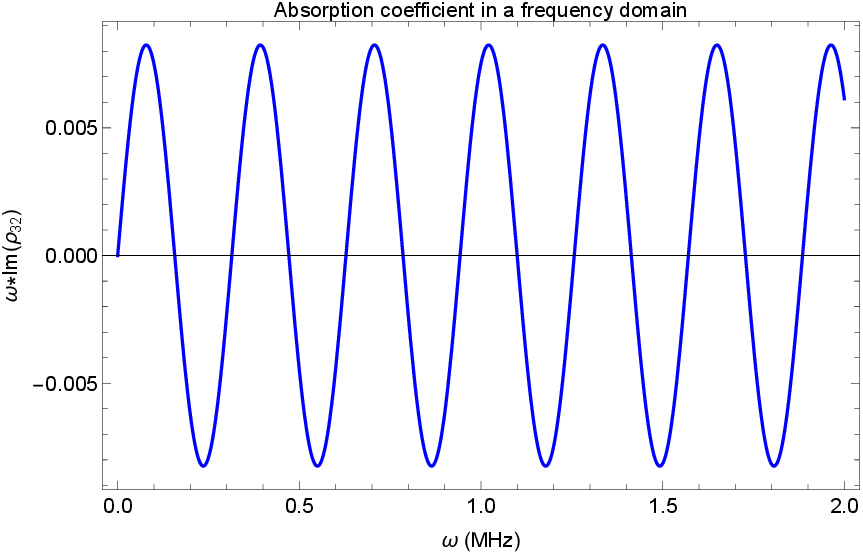}}
\et
\caption{
\textbf{A FREQUENCY MEASURE}: Absorption coefficient $\omega*\mathfrak{Im}(\omega)$ with magnetically-induced absorption ($> 0$) and transparency ($\leq 0$, where $< 0$ stands for stimulated emission (gain)) periodic ranges in a frequency domain for the parameters $\Omega = 0.1$ MHz, $\Gamma = 6.06536$ MHz, $\Delta_2 = 
0.7$ kHz, $L_g = 0.07$ MHz, $\rho_{11} (0) = 1/2$, $\mathfrak{Re}(\rho_{21}(0)) = 0$, $\mathfrak{Im}(\rho_{21}(0)) = 0$, $T = 20 \mu s$ (the time of Fourier transform or the time of fields' action) and in the quasi-stationary regime ($\dot{\rho}_{31} \approx \dot{\rho}_{32} \approx 0$).
}
\label{fig:AbsorpVLF}
\end{figure}

The characteristic absorption and dispersion responses of radio wave probes to ${}^{87}\text{Rb}$ vapor (or analogous systems) in a J-type configuration, under combined electromagnetic and magnetic driving fields, have slow oscillatory behavior (see Figs.~\ref{fig:AbsorpVLF}) [for very-low-frequency (VLF), it will be even slower].

Unlike the rapid oscillations in the visible range (see Figs.~\ref{fig:AbsorpVis} and~\ref{fig:DispVis}), radio-wave probes exhibit slow oscillatory behavior, making them suitable for weak magnetic-field detection.

%***************************************************************
%***************************************************************

%***************************************************************
\begin{figure}[t]
\bt{r}
\centerline{\includegraphics[width=1\columnwidth]{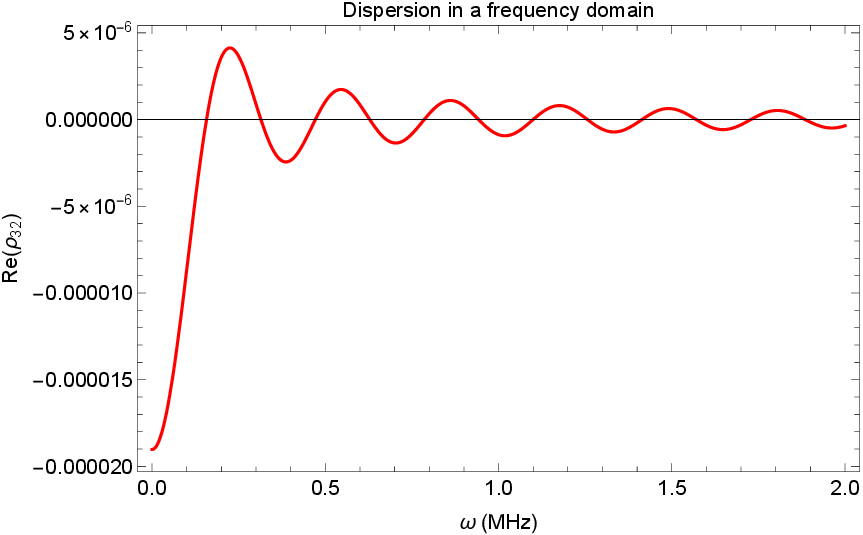}}
\et
\caption{
Dispersion $\mathfrak{Re}(\omega)$ with normal ($>0$) and anomalous ($<0$) periodic ranges in a frequency domain for the parameters  $\Omega = 0.1$ MHz, $\Gamma = 6.06536$ MHz, $\Delta_2 = 
0.7$ kHz, $L_g = 0.07$ MHz, $\rho_{11} (0) = 1/2$, $\mathfrak{Re}(\rho_{21}(0)) = 0$, $\mathfrak{Im}(\rho_{21}(0)) = 0$, $T = 20 \mu s$ (the time of Fourier transform or the time of fields' action) and in the quasi-stationary regime ($\dot{\rho}_{31} \approx \dot{\rho}_{32} \approx 0$).
}
\label{fig:DispVLF}
\end{figure}
%***************************************************************

\section{Vapor device as a Frequency Comb Generator}
\begin{figure}[t]
\bt{r}
\centerline{\includegraphics[width=1\columnwidth]{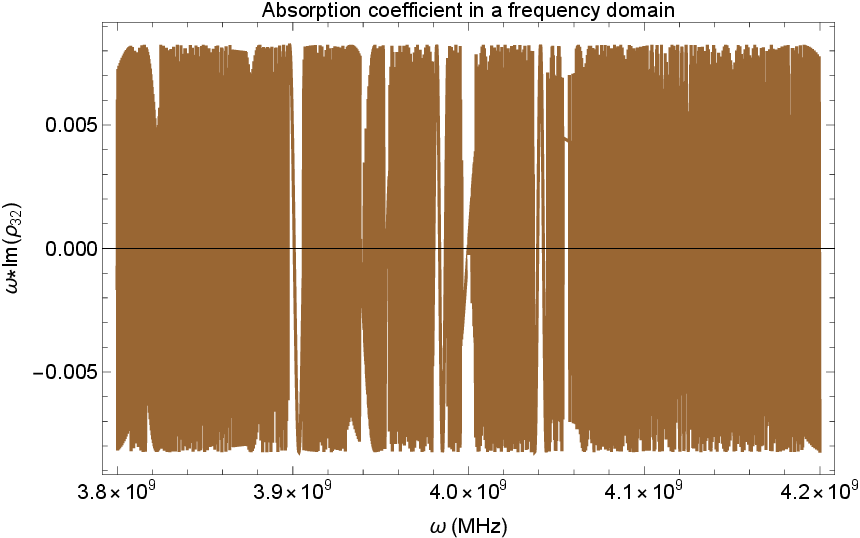}}
\et
\caption{
The case in Fig.~\ref{fig:AbsorpVLF}, but in the visible region. 
}
\label{fig:AbsorpVis}
\end{figure}

% Contrary, in the case of the visible light probe, a vapor device can act as a selective filter, i.e., it selectively absorbs one frequencies for which the absorption coefficient is positive and leave/intesify other frequencies for which the absorption coefficient is negative (MIT). Imagine that the incident probe white light with all visible continuum, the certain frequencies will be selectively transparent/intensified by the vapor device and, according to rapid oscillations in Fig.~\ref{fig:AbsorpVis}, it can be interesting candidate for a frequency comb generator. 
In contrast, for a visible-light probe, the vapor device can act as a selective spectral filter: it selectively absorbs frequencies for which the absorption coefficient is positive, while allowing or even enhancing frequencies for which the absorption coefficient is negative (MIT). If the incident probe is broadband white light covering the entire visible spectrum, certain frequencies will be preferentially transmitted or amplified by the vapor device. According to the rapid oscillatory behavior across the spectrum shown in Fig.~\ref{fig:AbsorpVis}, the system can filter a discrete set of spectral modes, being a promising candidate for a frequency comb generator.

\begin{figure}[t]
\bt{r}
\centerline{\includegraphics[width=1\columnwidth]{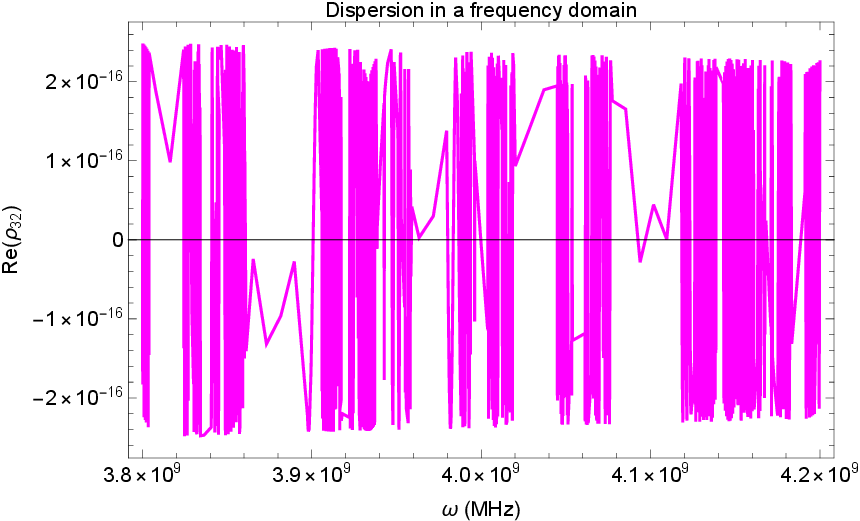}}
\et
\caption{
The case in Fig.~\ref{fig:DispVLF}, but in the visible region. 
}
\label{fig:DispVis}
\end{figure}

Rapid oscillatory behavior across the visible spectrum in the refractive-index difference (dispersion) corresponds to frequency regions where the medium alternates between normal and anomalous dispersion (see Fig.~\ref{fig:DispVis}). MINAD can produce spectrally selective group delays, and, in combination with MITA, it can enable frequency-comb–like filtering or shaping of the probe spectrum.

By controlling which frequencies are slowed (normal dispersion), one can temporarily store or delay optical pulses. Thus, MIT-based quantum memories can provide an alternative to EIT-based or other well-established approaches \cite{Sangouard2011, Simon2010, Tittel2009}. Contrary, the fast-light effects in anomalous dispersion region can have applications in temporal signal processing \cite{Li2015, Gui2020, Liao2021}, optical phase control, testing fundamental physics concepts like superluminal propagation \cite{Wang2000, Dogariu2001, Stenner2003}, etc \cite{Glasser2012, Goyon2021}. By engineering the frequency-dependent group delay, one can reshape pulses in time.

%***************************************************************
%***************************************************************

%***************************************************************
%***************************************************************

\section{Conclusions\label{Sec:concl}}

In this work, we have presented a comprehensive theoretical analysis of MITA and MINAD in a weakly excited ${}^{87}\text{Rb}$ atomic vapor, and more generally in a J-type three-level system, with a focus on applications to precision weak magnetic-field measurements. By accounting for both longitudinal and transverse components of the applied magnetic field and operating in the weak-excitation regime, we derived analytical solutions of the Bloch equations and obtained closed-form expressions for the relevant atomic populations and coherences in the stationary, quasi-stationary, and short-pulse regimes.

The separation of the system dynamics into two independent sets of differential equations enabled exact analytical treatment using eigenvector methods and the variation-of-constants technique, providing clear physical insight into the origin of the observed spectral features.

A key outcome of our analysis is the identification of a bifurcation in the oscillatory dynamics at zero longitudinal Zeeman splitting, which marks a transition from single-mode to dual-mode behavior. 

Explicit expressions for the absorption coefficient and dispersion, obtained from the imaginary and real parts of the coherence $\rho_{32}$, establish direct links between measurable optical responses and the underlying magnetic-field parameters. In the quasi-stationary regime, both MITA and MINAD exhibit oscillatory behavior in the Fourier domain, with frequency-dependent sign changes in absorption and refractive-index difference. These slow oscillations at radio-frequency probes make the system particularly attractive for sensitive detection of weak magnetic fields.

At higher probe frequencies, particularly in the visible domain, the oscillatory behavior of absorption and dispersion becomes rapid, and the atomic vapor effectively functions as a selective spectral filter and phase-shaping element, capable of extracting a broadband frequency comb from an incident optical continuum. This dual functionality---magnetometry at low frequencies and spectral filtering at optical frequencies---highlights the versatility of J-type atomic systems under combined electromagnetic and magnetic driving.

Overall, the results presented here provide a unified analytical framework for understanding MITA and MINAD across different dynamical regimes and frequency domains. They offer concrete guidance for experimental realization and optimization of atomic-vapor-based weak magnetic-field sensors, and open avenues for exploiting magnetically controlled dispersion and absorption in precision metrology and photonic signal processing.

%========================

%========================

%%%%%%%%%%%%%%%%%%%%%%%%%%%%%%%%%%%%%%%%%%%%%%%%%%%%%%%%%%%%%%%%%%%%%%%%%%%%%%%%%%%%%%%%%%%%%%%%%%%%%%%%%%%%%%%%%%%%%%%%%%%%%%%%%%%%%%%%%%%%%%%%%%%%%%%%%%%%%%%%

\acknowledgments
HLG acknowledges support from the Higher Education and Science Committee of Armenia in the frames of the research project 20TTAT-QTc004 on \textit{Quantum Technologies}, funded from 2021 to 2023. 

I would like to express my sincere gratitude to Prof. Dr. Yuri P. Malakyan for his supervision during the research grant and for introducing me to this fascinating topic, and to the grant manager Dr. Anahit Gogyan for her financial and broader support throughout the project.

%%%%%%%%%%%%%%%%%%%%%%%%%%%%%%%%%%%%%%%%%%%%%%%%%%%%%%%%%%%%%%%%%%%%%%%%%%%%%%%%%%%%%%%%%%%%%%%%%%%%%%%%%%%%%%%%%%%%%%%%%%%%%%%%%%%%%%%%%%%%%%%%%%%%%%%%%%%%%%%%
%%%%%%%%%%%%%%%%%%%%%%%%%%%%%%%%%%%%%%%%%%%%%%%%%%%%%%%%%%%%%%%%%%%%%%%%%%%%%%%%%%%%%%%%%%%%%%%%%%%%%%%%%%%%%%%%%%%%%%%%%%%%%%%%%%%%%%%%%%%%%%%%%%%%%%%%%%%%%%%%
%%%%%%%%%%%%%%%%%%%%%%%%%%%%%%%%%%%%%%%%%%%%%%%%%%%%%%%%%%%%%%%%%%%%%%%%%%%%%%%%%%%%%%%%%%%%%%%%%%%%%%%%%%%%%%%%%%%%%%%%%%%%%%%%%%%%%%%%%%%%%%%%%%%%%%%%%%%%%%%%

\end{document}